\documentclass[aps,letterpaper,preprint]{revtex4}
\usepackage{graphicx}
\usepackage{amsmath}
\usepackage{epstopdf}
\usepackage{amsfonts}
\usepackage{amssymb}
\usepackage{feynmf}

\begin{document}
\title{Conditions for extreme sensitivity of protein diffusion in membranes to cell environments}
\author{Yaroslav Tserkovnyak}
\affiliation{Lyman Laboratory of Physics, Harvard University, Cambridge, Massachusetts 02138, USA}
\affiliation{Department of Physics and Astronomy, University of California, Los Angeles, California 90095, USA}
\author{David R. Nelson}
\affiliation{Lyman Laboratory of Physics, Harvard University, Cambridge, Massachusetts 02138, USA}

\begin{abstract}
We study protein diffusion in multicomponent lipid membranes close to a rigid substrate separated by a layer of viscous fluid. The large-distance, long-time asymptotics for Brownian motion are calculated using a nonlinear stochastic Navier-Stokes equation including the effect of friction with the substrate. The advective nonlinearity, neglected in previous treatments, gives only a small correction to the renormalized viscosity and diffusion coefficient at room temperature. We find, however, that in realistic multicomponent lipid mixtures, close to a critical point for phase separation, protein diffusion acquires a strong power-law dependence on temperature and the distance to the substrate $H$, making it much more sensitive to cell environment, unlike the logarithmic dependence on $H$ and very small thermal correction away from the critical point.
\end{abstract}

\date{\today}
\maketitle

\begin{figure}
\includegraphics[width=0.9\linewidth,clip=]{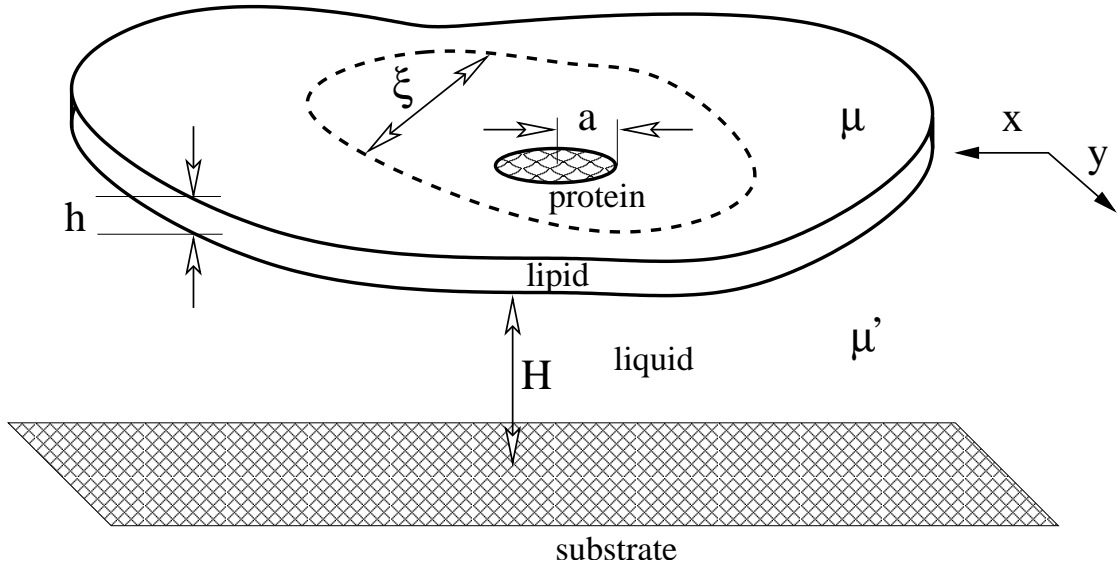}
\caption{Schematic of a flat two-dimensional lipid membrane of thickness $h$ separated by a layer of a bulk liquid of depth $H$ from a parallel rigid substrate. Since realistic lipid bilayers do not permit shear across the membrane, the velocity field in the membrane is taken to be strictly two-dimensional, characterized by a (two-dimensional) dynamic viscosity $\mu h$. The liquid is described by a dynamic viscosity $\mu^\prime$ (for water, $\mu^\prime\ll\mu$). No-slip boundary conditions are assumed at both the substrate and membrane interface. Two scenarios for thermally-driven protein diffusion are considered in the text: We first discuss Brownian motion of proteins of lateral size $a$ within a uniform lipid background. In the second model, proteins segregate in lipid ``rafts" of (predominantly) a particular chemical composition above the critical temperature $T_c$ of a binary lipid mixture. The large-scale diffusion behavior of proteins is then reformulated in terms of the raft diffusion of size $\xi$. Physically, $\xi$ is the correlation length of the binary mixture, which diverges as a power law near the critical temperature. As $\xi$ grows near $T_c$, protein diffusion becomes extremely sensitive to temperature and substrate depth $H$, see Eq.~(\ref{DTH}).}
\label{sc}
\end{figure}

We revisit here the problem of particle diffusion and relaxation of concentration fluctuations within a two-dimensional incompressible hydrodynamic medium \cite{saffmanPNAS75,saffmanJFM76} in proximity to a rigid substrate \cite{evansJFM88,stoneJFM98,barentinJFM99}. Interest in the two-dimensional Brownian motion has been largely motivated by protein diffusion in cell lipid membranes, with essential features captured by the seminal theory of Saffman and Delbr\"{u}ck \cite{saffmanPNAS75,saffmanJFM76} who predicted logarithmic dependence of the diffusion coefficient on a large-distance cutoff. The central result of the linearized theory \cite{saffmanPNAS75,saffmanJFM76} for the translational diffusion coefficient is
\begin{equation}
D\approx\frac{k_BT}{4\pi\mu h}\ln\frac{l^\ast}{a}\,,
\label{D}
\end{equation}
assuming an appropriate large-distance cutoff $l^\ast\gg a$, where $a$ is the small lateral size of the diffusing protein. See Fig.~\ref{sc} for a sketch of the physical situation. Here, $k_B$ is the Boltzmann's constant, $T$ the absolute temperature, $h$ the membrane thickness and $\mu$ its dynamic viscosity. The cutoff $l^\ast$ depends on membrane dimensions and also on environmental factors such as the viscosity of surrounding fluid or the distance to a nearby rigid wall. Eq.~(\ref{D}) allows an estimate of the protein size in terms of a known viscosity or vice versa, if one can selectively probe diffusion of a given protein.  Protein diffusion coefficients were measured, for example, by Peters and Cherry \cite{petersPNAS82} in large lipid vesicles  by monitoring fluorescence of diffusing proteins after a circular area of the membrane was illuminated by a laser. Owing to the logarithmic dependence of the diffusion coefficient (\ref{D}) on $a$, however, the precise form of the diffusion coefficient in terms of basic parameters of the problem is clearly required in order to make such techniques useful. Possible $l^\ast$ (and thus $D$) sensitivity to proximity to a rigid substrate can be used to investigate the physics and biology of membrane coupling to a nearby substrate such as a solid wall \cite{evansJFM88,stoneJFM98}. Proximity to a substrate affects transport and possibly also structural properties of the membrane, effects that could be biologically relevant for cell signaling and sensing of the external environment \cite{simonsARBBS04}. 

For an isolated flat two-dimensional membrane with lateral momentum conservation, $l^\ast$ is set simply by the membrane size. In practice, lateral momentum conservation requires a negligible viscosity of the surrounding fluid and an appreciable offset from any rigid walls. When $l^\ast\to\infty$, the diffusion coefficient (\ref{D}) diverges, which is reflected in the logarithmic factor in the mean-square displacement of a diffusing particle: $\langle r^2(t)\rangle\propto t\ln t$ \cite{saffmanPNAS75,saffmanJFM76}. In this case, the diffusion coefficient becomes effectively time dependent, growing logarithmically in time and eventually saturating at the value given by Eq.~(\ref{D}) with $l^\ast$ determined by geometric constraints (either a finite size or radius of curvature). The divergence of the linear mobility when $l^\ast\to\infty$ is the Stokes paradox \cite{lambBOOK32}: There is no solution of the low Reynold's number linearized-flow equations for steady translational motion in two dimensions. The time-dependent diffusion coefficient $D$ and mobility $b$ are however still related by the Einstein relation:
\begin{equation}
D=bk_BT\,.
\label{E}
\end{equation}
Upon taking into account the inertia of the viscous fluid (in the form of the advection), the nonlinear mobility becomes finite but logarithmically growing with the inverse velocity \cite{lambBOOK32}. A cutoff $l^\ast$ for the large-distance divergence of the diffusion coefficient and corresponding linear mobility related by Eq.~(\ref{E}) can be also provided by attaching the membrane to a momentum sink on either or both sides. Embedding the membrane into a slightly viscous three-dimensional liquid with viscosity $\mu^\prime\ll\mu$, for example, provides a cutoff $l^\ast\to R^\ast=(\mu/\mu^\prime)h$ \cite{saffmanPNAS75,saffmanJFM76}. For biological membranes with $\mu/\mu^\prime\sim100$, this is shorter than geometric cutoffs, provided we take $h\sim10$~nm and consider cells larger than $\mu$m.

\begin{figure}
\includegraphics[width=0.9\linewidth,clip=]{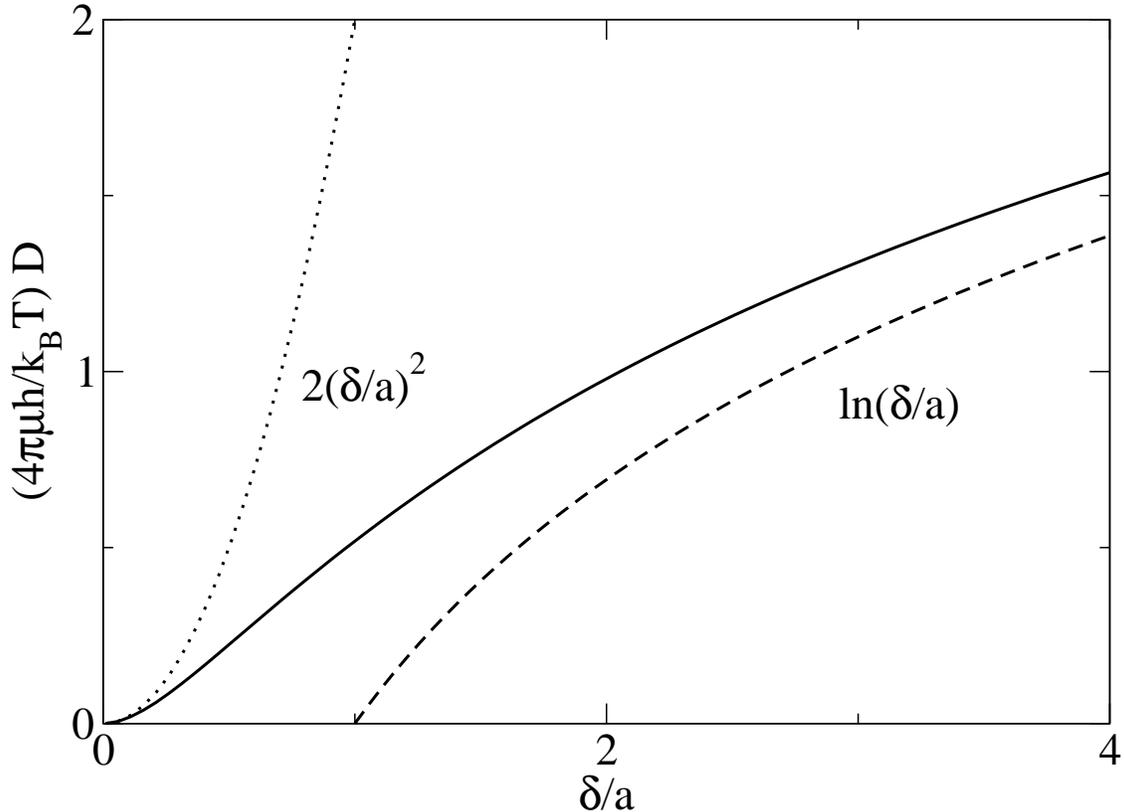}
\caption{The solid line shows the undimensionalized diffusion coefficient (\ref{ES}) in units of $k_BT/(4\pi\mu h)$ as a function of $\delta/a$. The dashed (dotted) line represents the large (small) $\delta/a$ limit discussed in the text. }
\label{Diff}
\end{figure}

Evans and Sackmann \cite{evansJFM88} discussed the problem of a coupling to a nearby rigid substrate, see Fig.~\ref{sc}. (Such a setup was used in Ref.~\cite{barentinJFM99} to build a two-dimensional shear viscometer.) When the coupling is mediated by a thin layer of viscous fluid of depth $H\ll R^\ast$ and viscosity $\mu^\prime$, the diffusion coefficient becomes
\begin{equation}
D=\frac{k_BT}{4\pi\mu h}\left[\frac{a}{\delta}\frac{K_1(a/\delta)}{K_0(a/\delta)}+\frac{a^2}{2\delta^2}\right]^{-1}\,,
\label{ES}
\end{equation}
where $K_0$ and $K_1$ are modified Bessel functions of the second kind of orders zero and one, respectively, and $\delta=\sqrt{R^\ast H}$. We plot Eq.~(\ref{ES}) in Fig.~\ref{Diff}. When $\delta\gg a$, Eq.~(\ref{ES}) reduces to the form (\ref{D}) with $l^\ast=\delta$. Increasing $H$ naturally increases the relevant cutoff $\delta(H)$; however, the assumptions leading to Eq.~(\ref{ES}) break down when $H\gtrsim R^\ast$ \cite{stoneJFM98} and the cutoff $l^\ast$ saturates at $\delta(H=R^\ast)=R^\ast$, where the remote rigid substrate becomes unimportant. In the opposite limit of $\delta\ll a$ (while still $H\ll R^\ast$), Eq.~(\ref{ES}) reduces to $D\approx k_B TH/(2\pi a^2\mu^\prime)$, from which one easily recognizes the mobility determined by the direct laminar-flow friction of the particle with the substrate. In these and similar linearized low-temperature theories \cite{stonePF95,stoneJFM98,barentinJFM99}, flow equations are solved for specific geometries. For example, the diffusing particle can be assumed to be a rigid cylinder of height $h$ and radius $a$ (with, e.g., a no-slip or no-tangential-stress boundary condition for liquid flow \cite{saffmanPNAS75,saffmanJFM76}). In the following, we present an approach that focuses on the long-time large-distance physics from the onset, without reference to any short-distance details, apart from a characteristic protein size $a$.

Our interest in the problem is twofold. On the one hand, we want to understand the interplay between thermal fluctuations and advective nonlinearity omitted in previous treatments of protein diffusion and linear mobility. In addition, we wish to investigate the behavior near a critical temperature $T_c$ of the multicomponent lipid mixtures that make up biological membranes. One can efficiently deal with nonlinearities using the dynamic renormalization-group (RG) approach \cite{maPRB75,forsterPRA77,aronovitzPRA84, hohenbergRMP77}, which will allow us to scale the large-distance long-time physics onto the ``asymptotically free" (i.e., linear) theory. We will then apply some general ideas from the theory of dynamic critical phenomena \cite{hohenbergRMP77} to the lipid and protein diffusion close to a critical point. Of a central interest to this paper is the sensitivity of these effects to proximity of a rigid wall, which exerts friction and bleeds momentum from the membrane. We thus want to generalize the result (\ref{ES}) to include the interplay between nonlinear corrections and thermal noise and to also adapt the theory to include critical fluctuations in multicomponent lipid membranes.

Before presenting details of the model and calculation, let us first outline what we can anticipate based on some simple but general arguments \cite{hohenbergRMP77}. Suppose the binary lipid membrane is close to a critical point for phase separation. Below the critical temperature $T_c$, lipids form laterally--phase-separated regions (``rafts"). Specialized proteins can segregate in lipid rafts for cell signaling, protein organization, and endocytosis. Raft-like structures were also observed in artificial lipid membranes consisting of a binary mixture with one phase rich in the unsaturated lipids and the other rich in the saturated lipids and cholesterol \cite{veatchBPJ03,baumgartNAT03,baumgartBPJ05}. Below $T_c$, phase-separated membranes also have inhomogeneous elastic properties, which lead to formation of nontrivial spatial structures \cite{baumgartNAT03,baumgartBPJ05} that are beyond our analysis. For simplicity, we focus on membrane diffusion at temperatures slightly above $T_c$, where lipids form nearly uniform mixture with correlation length diverging on approaching $T_c$. If proteins strongly prefer one kind of lipid versus the other, they will preferentially segregate in the correlated regions (which we can also consider as ``rafts") corresponding to one of the emerging phases. On time scales longer than it takes for the protein to diffuse across such rafts, the overall protein motion becomes essentially passive and governed by lipid concentration intermixing. It will be shown that the protein size $a$ must then be replaced by the correlation length $\xi$ in order to obtain the effective diffusion coefficient. The latter is then approximately given by the Einstein relation (\ref{E}) with mobility $b$ scaling as $\xi^{2-d}/\mu$ (assuming momentum conservation within the membrane), which is inversely proportional to the drag force exerted by the membrane fluid on a drifting raft of size $\xi$, where we in general consider a $d$-dimensional system (eventually setting $d=2$). We thus have
\begin{equation}
D\mu\propto k_BT\xi^{2-d}\,,
\label{Dmu}
\end{equation}
where in general both $D$ and $\mu$ can be temperature dependent. For $d>2$, the right-hand side of Eq.~(\ref{Dmu}) vanishes near the critical temperature where $\xi$ diverges. Short-range physics on length scales shorter than $\xi$, which is disregarded here, then has to be invoked in order to calculate the product $D\mu$. When $d<2$, on the other hand, Eq.~(\ref{Dmu}) diverges near $T_c$ signaling emergence of important long-distance processes. $d=2$ is the critical dimension where one may expect only logarithmic singularities due to long-distance physics, as we will indeed see in the following. When there is a rigid substrate exerting a friction on the two-dimensional membrane dynamics, however, the diffusion coefficient can acquire a power-law dependence on both $\xi$ and the distance to the substrate, making the diffusion very sensitive to the environment near $T_c$. In particular, as explained in the following, there is a regime where the protein size $a$ is effectively replaced by the correlation length $\xi$ in Eq.~(\ref{ES}), resulting in Eq.~(\ref{DTH}) for large $\xi$. Note that in this case $D$ depends on the viscosity of the surrounding fluid and not the membrane viscosity itself.

Eq.~(\ref{Dmu}) is also relevant for diffusion away from the critical point, where $\xi$ may be replaced by $1/q$ and the diffusion coefficient and viscosity become wavevector-dependent functions, $D(q)$ and $\mu(q)$. The long-distance physics at $q\to0$ is important for $d\leq2$, consistent with the result of the linearized theory for $d=2$, which gives
\begin{equation}
D\mu\propto\ln(1/qa)
\label{qa}
\end{equation}
for wavelengths shorter than $l^\ast$ [obtained by replacing $l^\ast$ with $1/q$ in Eq.~(\ref{D})]. We will show that although nonlinearities and thermal fluctuations leave Eq.~(\ref{qa}) unmodified away from the critical point, the logarithmic singularity is \textit{not} entirely contained in $D$; in fact, the effective viscosity $\mu$ may also acquire a finite-temperature logarithmic $q$ dependence. We study this phenomenon first, before discussing diffusion near the critical point, in order to understand how the membrane viscosity itself depends on temperature and relevant length scales.

Let us now formulate the model. Focusing on the long-wavelength two-dimensional dynamics, we write Navier-Stokes equation for membrane fluid velocity field subject to a random forcing function $\mathbf{f}(\mathbf{r},t)$:
\begin{equation}
\partial_t\mathbf{v}+\lambda(\mathbf{v}\cdot\boldsymbol{\nabla})\mathbf{v}=-\frac{1}{\rho}\boldsymbol{\nabla}p-\alpha\mathbf{v}+\nu\nabla^2\mathbf{v}+\mathbf{f}\,,
\label{ns}
\end{equation}
where all vectors have two components in the $xy$ plane, e.g., $\mathbf{r}=(x,y)$. The field $\mathbf{v}(\mathbf{r},t)$ is the position- and time-dependent fluid velocity, $\rho$ is a constant mass density and $\nu=\mu/\rho$ the kinematic viscosity of the membrane fluid. (In writing a strictly two-dimensional Navier-Stokes equation, we have assumed a large planar shear viscosity within the membrane, relative to the fluid outside.) The coupling $\alpha$ describes friction due to a nearby rigid substrate. If the friction is mediated by a constant-gradient laminar flow of a liquid layer of thickness $H$ with viscosity $\mu^\prime$, then $\alpha=\mu^\prime/(hH\rho)$. (For simplicity, we introduce the coupling $\alpha$ only on one side of the membrane.) We are assuming here that $R^\ast=(\mu/\mu^\prime)h\gg H$; otherwise the membrane does not sense the substrate and $H$ has to be replaced by $R^\ast$ in the definition of $\alpha$ for lateral lengthscales smaller than $R^\ast$. The pressure term $-(1/\rho)\boldsymbol{\nabla}p$ enforces incompressible fluid flow:
\begin{equation}
\boldsymbol{\nabla}\cdot\mathbf{v}=0\,.
\label{ic}
\end{equation}
The second term on the l.h.s. of Eq.~(\ref{ns}) is the usual (nonlinear) advective term when $\lambda=1$. We will be able to treat it using an RG approach \cite{forsterPRA77} where $\lambda$ becomes a perturbative parameter weakly irrelevant for large-distance asymptotics. The fluctuation-dissipation theorem dictates the form of $\mathbf{f}$ in correspondence with the dissipative terms in Eq.~(\ref{ns}):
\begin{equation}
\langle f_i(\mathbf{k},\omega)f_j(\mathbf{k}^\prime,\omega^\prime)\rangle=2\tau(\alpha+\nu k^2)P_{ij}(\mathbf{k})(2\pi)^3\delta(\mathbf{k}+\mathbf{k}^\prime)\delta(\omega+\omega^\prime)\,,
\label{fdt}
\end{equation}
where $\mathbf{f}(\mathbf{k},\omega)=\int d^2\mathbf{r}dt\mathbf{f}(\mathbf{r},t)\exp(i\omega t-i\mathbf{k}\cdot\mathbf{r})$ is the Fourier transform of the velocity distribution, $\tau=k_B T/(h\rho)$ is a rescaled temperature, and $P_{ij}(\mathbf{k})=\delta_{ij}-k_ik_j/k^2$ is the transverse projection operator. We note that when coupling $\alpha$ is mediated by a thin liquid interlayer connecting the membrane with the substrate, we account for it by assuming a simple laminar flow, which is not perturbed by thermal fluctuations and which responds instantaneously to the driving shear provided by the membrane dynamics. The nonlinearities and stochasticity are fully contained within the membrane. We thus neglect quasi-two-dimensional nonlinear and thermal effects as well as inertia in the interlayer. However, we expect that these complications simply lead to a renormalization of parameters such as $\lambda$, $\rho$, $\alpha$, and $\nu$ in Eq.~(\ref{ns}).

The particle (protein) concentration $c$ is treated in the dilute limit as a passive scalar characterized by some local diffusion coefficient $D_0$ and statistically advecting along fluid flows (with no back-reaction on the hydrodynamics) described by Eqs.~(\ref{ns})-(\ref{fdt}):
\begin{equation}
\partial_t c+\lambda(\mathbf{v}\cdot\boldsymbol{\nabla})c=D_0\nabla^2c\,,
\label{pd}
\end{equation}
which can be thought of as the Fokker-Planck equation for the probability distribution of an individual particle undergoing Brownian motion \cite{forsterPRA77}. At this point, we disregard protein coupling to the rigid substrate and assume that it is felt only indirectly through the fluid friction $\alpha$ and particle's drag by the fluid flow. Therefore, $\lambda=1$ in Eq.~(\ref{pd}). As explained in the Supplemental Material, this coupling then obeys the same RG flows upon coarse graining as $\lambda$ in Eq.~(\ref{ns}): that is why we are using the same variable $\lambda$ in the two cases. For sufficiently shallow depths $H$, however, the fluid-mediated particle-substrate friction becomes important, as already mentioned in the linearized regime described by Eq.~(\ref{ES}). This can be easily taken into account in the final result for the mobility [and thus diffusion coefficient, according to the Einstein relation (\ref{E})], as an additional drag by the rigid substrate on the diffusing particle. In the following, we solve Eqs.~(\ref{ns})-(\ref{pd}), focusing on the viscosity and diffusion-constant renormalization in the infrared limit.

Let us first calculate viscosity renormalization due to the advection term in Eq.~(\ref{ns}). Proceeding along the lines of Ref.~\cite{forsterPRA77}, we Fourier transform Eq.~(\ref{ns}) to obtain
\begin{equation}
v_l(\mathbf{k},\omega)=G_0(k,\omega)f_l(\mathbf{k},\omega)-\frac{i}{2}\lambda P_{lij}(\mathbf{k})G_0(k,\omega)\int\frac{d^2\mathbf{q}d\Omega}{(2\pi)^3}v_{i}(\mathbf{q},\Omega)v_{j}(\mathbf{k}-\mathbf{q},\omega-\Omega)\,,
\label{vl}
\end{equation}
where $P_{lij}(\mathbf{k})=k_{i}P_{jl}(\mathbf{k})+k_{j}P_{il}(\mathbf{k})$, $G_{0}(k,\omega)=(-i\omega+\alpha+\nu k^{2})^{-1}$ is the response function of the linear theory with $\lambda=0$, and summation over repeated indices is implied. Eq.~(\ref{vl}) can be solved perturbatively in $\lambda$ by iteration. We first introduce a short-distance cutoff $1/\Lambda$ for the velocity field $\mathbf{v}$ and the stochastic field $\mathbf{f}$. Physically, $1/\Lambda$ is of the order of the interlipid separation, the scale at which the continuous hydrodynamics obviously fails. Correspondingly, the Fourier transforms must be cut off at momentum $\Lambda$. The lowest-order correction to the full response function $G(k,\omega)$ due to the advection term $\propto\lambda$ results in the following contribution to the wave-vector--dependent viscosity (setting $\lambda=1$) \cite{forsterPRA77}:
\begin{equation}
\Delta\nu\approx\frac{\tau}{16\pi\nu}\ln(\Lambda/k)
\label{Dnu}
\end{equation}
at $\omega=0$, assuming $k$ is larger than the momentum cutoff $1/\delta=\sqrt{\nu/\alpha}=\sqrt{(\mu/\mu^\prime)hH}$ set by a finite $\alpha$. When $k\ll1/\delta$, the argument of the logarithm in Eq.~(\ref{Dnu}) must be replaced by $\Lambda\delta$. The importance of advection due to thermal fluctuations described by Eq.~(\ref{Dnu}) thus hinges on the size of the prefactor $\tau/(16\pi\nu)$. Taking room temperature, water density, and $h\sim10$~nm for the evaluation of $\tau=k_BT/(h\rho)$ and setting membrane viscosity $\nu$ to be 100 times larger than that of water, we get $\tau/(8\pi\nu^{2})\sim10^{-9}$. Clearly this gives only a tiny correction to the linearized theory for biologically-relevant problems.

It is nevertheless interesting to pursue a nonperturbative calculation in the infrared limit $k\to0$ for a very small damping $\alpha\to0$ leading to a logarithmic divergence of the correction (\ref{Dnu}). This exercise allows additional insights into the peculiarities of the two-dimensional dynamics. In order to discuss the nonperturbative asymptotics, we proceed using the dynamic RG \cite{maPRB75,forsterPRA77} in two standard steps: First, we reduce the cutoff to $\Lambda e^{-l}$, where $l\ll1$, and rescale parameters $\alpha$, $\nu$, and $\lambda$ of the Navier-Stokes equation (\ref{ns}). The requirement is that going from initial (physical) set of parameters $(\Lambda,\alpha,\nu,\lambda)$ to $[\Lambda e^{-l},\alpha(l),\nu(l),\lambda(l)]$ does not modify velocity correlators and response functions averaged over the random force (\ref{fdt}). Second, we rescale momentum by $e^{l}$, so that the cutoff is unchanged by the two steps, and also conveniently choose frequency rescaling. The velocity field is rescaled by requiring that the relation (\ref{fdt}) remains unmodified in terms of the $l$-dependent $\alpha$ and $\nu$. Finally, the parameters $\alpha$, $\nu$, and $\lambda$ are correspondingly renormalized. This procedure is then repeated continuously with accumulated rescaling exponent $l$ going from 0 to a chosen value much greater than 1.

Since we adapted the scheme developed in Refs.~\cite{maPRB75,forsterPRA77}, we will avoid repeating the details of the derivation. A summary is given in the Supplemental Material. Solving Eq.~(\ref{dn}) in the Supplemental Material gives for the renormalized viscosity at $k\to0$
\begin{equation}
\nu_{r}\approx\sqrt{\nu^{2}+\frac{\tau}{8\pi}\ln(\Lambda\delta_{r})}\,,
\label{nus}
\end{equation}
where
\begin{equation}
\delta_{r}=\sqrt{\frac{\mu_{r}}{\mu^{\prime}}hH}
\label{dr}
\end{equation}
itself depends on the renormalized (dynamic) viscosity $\mu_{r}=\nu_{r}\rho$. If the membrane is nearly isolated, i.e., $\alpha=\mu^\prime/(hH\rho)\approx0$, the second summand under the square root in Eq.~(\ref{nus}) dominates and the renormalized viscosity grows logarithmically when $k\to0$, eventually saturating at the value determined by the distance cutoff $\delta_r$:
\begin{equation}
\nu_{r}(k\to0)\approx\sqrt{\frac{\tau}{8\pi}\ln\left(\frac{\Lambda}{k}\right)}\to\sqrt{\frac{\tau}{8\pi}\ln\left(\Lambda\delta_r\right)}\,,
\label{na}
\end{equation}
which agrees with the result obtained in Ref.~\cite{forsterPRA77}. In practice, however, as we have discussed, this limit is hard to achieve. The thermal renormalization of the viscosity due to the advective term in the Navier-Stokes equation (\ref{ns}) turns out to be very small and instead $\nu_r\approx\nu$.

Having studied the large-scale long-time velocity-field dynamics, we will now discuss consequences for the passive-scalar advection described by Eq.~(\ref{pd}). This equation can be solved iteratively treating $\lambda$ as a perturbative parameter, similarly to the velocity-field expansion (\ref{vl}) but with a slightly different physical formulation: For the velocity field $\mathbf{v}(\mathbf{r},t)$, we were interested in the response to thermal and external forces driving the fluid dynamics; for the particle diffusion, on the other hand, we do not have any particle sources or sinks, but instead want to consider the evolution of the initial distribution $c(\mathbf{r},t=0)$ at $t>0$. It is thus natural to Fourier transform $c(\mathbf{r},t)$ in space but Laplace transform in time: $c(\mathbf{k},\omega)=\int d^2\mathbf{r}\int_0^\infty dt\exp(i\omega t-i\mathbf{k}\cdot\mathbf{r})$, where $\omega$ has an infinitesimal positive imaginary component. The scalar field $c(\mathbf{k},\omega)$ is then given according to Eq.~(\ref{pd}) by
\begin{equation}
c(\mathbf{k},\omega)=\mathcal{G}_0(k,\omega)c(\mathbf{k},t=0)-i\lambda\mathcal{G}_0(k,\omega)k_i\int\frac{d^2\mathbf{q}d\Omega}{(2\pi)^3}c(\mathbf{q},\Omega)v_i(\mathbf{k}-\mathbf{q},\omega-\Omega)\,,
\label{c}
\end{equation}
where $\mathcal{G}_0(k,\omega)=(-i\omega+D_0k^2)^{-1}$ is the diffusion propagator [$\mathbf{v}(\mathbf{k},\omega)$ is still the full Fourier transform, as before]. The technical details for solving Eq.~(\ref{c}) are once again summarized in the Supplemental Material. Interestingly, we find that $D_r\sim\ln^{1/2}(\delta_r/a)$ at the very longest lengths due to thermal fluctuations, in contrast to Eq.~(\ref{D}). However, at the temperatures of interest for membranes, we find in practice that $\nu_r\approx\nu$ and $\nu_r\gg D_r$. Eq.~(\ref{dD}) of the Supplemental Material then gives for $k\to0$ simply
\begin{equation}
D_r\approx \frac{\tau}{4\pi\nu}\ln\left(\frac{\delta_r}{a}\right)\,,
\label{Dr}
\end{equation}
neglecting the short-distance contribution $D_0$ and assuming $\delta_r\gg a$. This is exactly the Saffman-Delbr{\"{u}}ck result~(\ref{D}) for the relevant distance cutoff $l^\ast=\delta_r$. We furthermore make the ansatz that at long times and distances for the diffusion problem one can in general use the results of the linearized theory, such as Eqs.~(\ref{D}) and (\ref{ES}), as long as the renormalized viscosity $\nu_r$ and self-consistently--found distance cutoff $\delta_r$ [Eq.~(\ref{dr})] are used. We expect this to be correct, up to numeric factors of order unity (such as the factor $2/A$ in the Supplemental Material), even in the extreme high-temperature limit when thermal fluctuations dominate equilibrium properties. The final result for the renormalized diffusion constant in the system sketched in Fig.~\ref{sc} is thus a trivial generalization of Eq.~(\ref{ES}):
\begin{equation}
D_r\approx\frac{k_BT}{4\pi\mu_rh}\left[\frac{a}{\delta_r}\frac{K_1(a/\delta_r)}{K_0(a/\delta_r)}+\frac{a^2}{2\delta_r^2}\right]^{-1}\,.
\label{ESr}
\end{equation}
Like Eq.~(\ref{ES}), this result includes the fluid-mediated friction between protein and the substrate, which we have not explicitly considered in the above discussion. As we have already explained, we can neglect thermal renormalization of $\nu_r$ (and thus $\delta_r$ and $\mu_r$) for biological systems.

We now discuss how these results are modified close to a critical point for phase separation in binary lipid membranes, leading to a much more dramatic dependence of diffusion on temperature and membrane environment. To be specific, let us consider concentration interdiffusion in a symmetric binary mixture of, say, lipids $A$ and $B$ with concentrations $c_A(\mathbf{r},t)$ and $c_B(\mathbf{r},t)$ \cite{hohenbergRMP77}. By symmetry, the critical point corresponds to equal average concentrations: $\langle c_A\rangle=\langle c_B\rangle=c/2$, where $c$ is the averaged total concentration (taken to be constant). The order parameter $\psi(\mathbf{r},t)$ is given by the normalized difference of the concentrations: $\psi(\mathbf{r},t)=[c_A(\mathbf{r},t)-c_B(\mathbf{r},t)]/c$. The total concentration, $c_A(\mathbf{r},t)+c_B(\mathbf{r},t)$, does not couple to the $\psi$ diffusion mode in this case. Following Ref.~\cite{hohenbergRMP77}, we introduce a fictitious external potential $\Phi$, which couples to $\psi$, by choosing opposite potentials for the two lipid components: $\mu_A=\Phi$ and $\mu_B=-\Phi$. The force on a region with a roughly uniform order parameter $\bar{\psi}$ of size $\xi$ (the order-parameter correlation length) is then given by 
\begin{equation}
\mathbf{f}_\Phi=-\boldsymbol{\nabla}F=-\xi^2\bar{\psi}\boldsymbol{\nabla}\Phi\,,
\end{equation}
assuming a long-wavelength (on the scale of $\xi$) driving field $\Phi$, where $F$ is the free energy. The region will accelerate until reaching velocity $\mathbf{v}$ when $\mathbf{f}_\Phi$ is canceled by the viscous drag $\mathbf{f}_\nu=-\mathbf{v}/b$ characterized by mobility $b$: $\xi^2\bar{\psi}\boldsymbol{\nabla}\Phi+\mathbf{v}/b=0$. We thus get for the net current density of the order parameter $\psi$ within the region of size $\xi$:
\begin{equation}
\mathbf{j}_\psi=\bar{\psi}\mathbf{v}=-b\xi^2\bar{\psi}^2\boldsymbol{\nabla}\Phi=-\sigma\boldsymbol{\nabla}\Phi\,,
\label{jp}
\end{equation}
where
\begin{equation}
\sigma=b\xi^2\bar{\psi}^2=bk_BT\chi
\label{lb}
\end{equation}
is the concentration conductivity associated with the potential $\Phi$ and $\chi$ is the susceptibility of the concentration difference $\psi$. The last equality in Eq.~(\ref{lb}) is obtained by invoking the fluctuation-dissipation theorem,
\begin{equation}
\langle\bar{\psi}^2\rangle=(k_BT/\xi^2)\chi\,,
\label{fdt2}
\end{equation}
implying averaging over correlated regions (over the entire system). For a smooth static external potential, $\psi=-\chi\Phi$ in equilibrium, and thus the current density $\mathbf{j}_\psi$ must be proportional to $\boldsymbol{\nabla}(\Phi+\psi/\chi)$. In the absence of the fictitious potential $\Phi$, the concentration flows are then described by the diffusion coefficient $D$: $\mathbf{j}_\psi=-D\boldsymbol{\nabla}\psi$ with
\begin{equation}
D=\frac{\sigma}{\chi}=bk_BT\,,
\label{Exi}
\end{equation}
where we have used Eq.~(\ref{lb}) for the last equality. Since $\psi$ is a conserved quantity, $\partial_t\psi=-\boldsymbol{\nabla}\cdot\mathbf{j}_\psi=D\nabla^2\psi$. Eq.~(\ref{Exi}) is simply the Einstein relation (\ref{E}), with one conceptual subtlety: $D$ and $b$ describe not the transport coefficients of individual small particles, but rather the (inter)diffusion coefficient of a binary mixture and the mobility of a region of size $\xi$, respectively. In particular, we can approximate the latter by the mobility of a ``particle" of size $\xi$, which in turn corresponds to the diffusion coefficient (\ref{ESr}). We conclude from this line of argument that the diffusion coefficient of proteins ``riding" rafts with order parameter $\bar{\psi}$ of a given sign [determined by protein affinity to a given kind of lipids, either $A$ ($\bar{\psi}>0$) or $B$ ($\bar{\psi}<0$)] is given by equation (\ref{ESr}) with $a\to\xi$. In particular, as $T\to T_c$ and the correlation length $\xi$ diverges, $\xi\gg\delta_r$ and the diffusion coefficient approaches
\begin{equation}
D\approx\frac{k_B TH}{2\pi \xi^2\mu^\prime}
\label{DTH}
\end{equation}
(assuming $H\ll R^\ast$), as we discussed in the introduction (see also the small-$\delta$ asymptotic of Fig.~\ref{Diff}). The membrane viscosity $\mu$ in fact also has a diverging contribution near $T_c$, but fortunately with a very small exponent ($\approx0.16$ in two dimensions) \cite{hohenbergRMP77}, so that as long as $\xi\gg\delta_r$ it drops out of Eq.~(\ref{DTH}) for the diffusion coefficient. The exponents that govern the correlation-length and susceptibility
\begin{equation}
\chi(T)\propto\xi(T)^{2-\eta}
\label{chi}
\end{equation}
divergence near the critical temperature can be calculated in the two-dimensional Ising model (which is in the same universality class as the binary mixture), giving $\xi\propto(T-T_c)^{-1}$ and $\eta=1/4$. Upon combining Eqs.~(\ref{Exi}), (\ref{DTH}), and (\ref{chi}), we find for the conductivity scaling $\sigma(T)\propto\xi(T)^{-\eta}$, which vanishes with a small exponent $\eta$ at the critical temperature. In practice, however, equation (\ref{lb}) with $b$ standing for the mobility of a rigid region of size $\xi$ neglects short-range (on the scale of $\xi$) fluctuations that can lead to a finite conductivity $\sigma_0$ at the critical temperature. The diffusion coefficient (\ref{Exi}) then has to be correspondingly corrected, giving near $T_c$ (with $\xi\gg\delta_r$)
\begin{equation}
D\approx\frac{k_B TH}{2\pi \xi^2\mu^\prime}+\frac{\sigma_0}{\chi_0}\left(\frac{a_0}{\xi}\right)^{2-\eta}\,,
\label{Df}
\end{equation}
with the second contribution eventually dominating very near $T_c$. The constants $\sigma_0$ and $\chi_0$ are nonuniversal, dependent on nonuniversal details of the lipid mixtures; $a_0$ is a characteristic lipid spacing.

We conclude with a discussion of one final point: Although the correlation length grows, density fluctuations eventually disappear at the critical temperature and the protein diffusion is not critically slowed down. Upon combining Eqs.~(\ref{fdt2}) and (\ref{chi}), we get $\langle\bar{\psi}^2\rangle=C(a_0/\xi)^\eta$ near $T_c$, in terms of some dimensionless constant $C$. We make a mean-field approximation for the free-energy difference for a single protein residing in region $A$ ($\bar{\psi}>0$) and region $B$ ($\bar{\psi}<0$), $\delta F\approx\kappa\delta\bar{\psi}/2$, where $\delta\bar{\psi}$ is the order-parameter difference between the regions, and assume that the proteins will thermally hop between the correlated regions, and thus freely diffuse within the membrane when $k_BT\sim\kappa\delta\bar{\psi}/2\sim\kappa\sqrt{C}(a_0/\xi)^{\eta/2}$, i.e., when the correlation length becomes larger than
\begin{equation}
\xi_{\rm max}\sim a_0\left(\frac{\kappa\sqrt{C}}{k_BT}\right)^{2/\eta}\,,
\label{xi}
\end{equation}
a relation which in two dimensions is characterized by a large exponent $2/\eta=8$. \textit{Very} close to the critical point where the correlation length is larger than Eq.~(\ref{xi}), the protein diffusion is thus not profoundly affected by the critical behavior apart from the weak singularity in the membrane viscosity.

We can make a crude estimate of $\xi_{\rm max}$ as follows. Suppose $\kappa$ is of order       
the energy difference between a typical protein embedded in the two-chain lipids and a typical protein embedded in the one-chain lipids. A typical protein size is 50~\AA; with a lipid head size of 6-10~\AA, maybe 20 lipids (ten from each leaflet) surround each protein. Assume the protein spans the bilayer, so each of $\sim12$ CH$_2$ groups of the lipid chains above and below can interact with the protein. A typical hydrogen bond or van~der~Waals attractive interaction with a CH$_2$ might be $0.2k_BT$ at room temperature. Since roughly 100 CH$_2$ groups are in contact in one case (single hydrocarbon chain), and 200 CH$_2$ groups in the other (two hydrocarbon chains), the energy difference is $\kappa\sim20k_BT$. In situations where both lipid species have two hydrocarbon chains, the energy difference could arise from the different head groups (not included in the above) and from the typical fact that one species has an unsaturated bond in the hydrocarbon chain (making protein interactions          
with part of the chain difficult) and the other does not.  Also, the chain lengths might differ by 2-3 CH$_2$ groups. All these factors together might lead to
\begin{equation}
\kappa\sim10k_BT
\label{kappa}
\end{equation}
in less favorable situations. The constant $C$ can be estimated by the mean-field approximation of the Ising model \cite{goldenfeldBOOK92}: Using the form $L=at\psi^2+b\psi^4/2+\gamma(\nabla\psi)^2/2$ [$t=(T-T_c)/T_c$ is the reduced temperature] of the Landau free-energy density, one gets for the mean-field response function $\chi=\xi^2/\gamma$ and the correlation length $\xi=\sqrt{\gamma/2at}$. According to Eq.~(\ref{DTH}) then $C=k_BT/\gamma$ and $\eta=0$ (instead of the correct $\eta=1/4$ since the mean-field approximation captures the exponents correctly only at more than four dimensions). For the Ising model, a mean-field estimate gives
\begin{equation}
C=k_BT/\gamma\sim(a_0/R)^2\sim1
\label{C}
\end{equation}
for short-range interactions, where $R$ is the range of exchange coupling. Using the estimates (\ref{kappa}) and (\ref{C}), we finally conclude that $\xi_{\rm max}\sim10^8a_0$ according to Eq.~(\ref{xi}), which clearly suggests that the reasoning leading to Eq.~(\ref{Df}) holds for a large range of the correlation length $\xi$ (apart from temperatures \textit{very} close to $T_c$ where the critical slowing down is suppressed).

In summary, we have considered the effects of thermal noise, hydrodynamic nonlinearities, and proximity to a critical point on protein diffusion in multicomponent lipid membranes close to a rigid substrate. We have found that nonlinear advection in the Navier-Stokes equation in practice gives only a small correction to the Evans-Sackmann result (\ref{ES}) for a uniform lipid composition. The protein diffusion is then only logarithmically sensitive to the distance to the wall, assuming the protein size $a\ll\delta=\sqrt{(\mu/\mu^\prime)hH}$. Upon approaching the critical temperature for the phase separation in multicomponent membranes, however, the long-time asymptotic of protein diffusion can be reformulated in terms of the lipid ``raft" diffusion of size $\xi$, which diverges at $T_c$. The diffusion coefficient then becomes \textit{linearly} sensitive to the distance to the substrate $H$ and has a roughly $1/\xi^2$ scaling with the correlation length, near $T_c$ (apart from a narrow temperature window very close to $T_c$ where proteins thermally hop between the rafts). If it were functionally advantageous for cell sensing and signaling, one might imagine that lipid composition could evolve to be near a critical point.

\acknowledgments

We thank A.~Ajdari, B.~I. Halperin, and H.~A. Stone for critical reading of the manuscript and valuable discussions. D.R.N. thanks K.~Binder and P.~C. Hohenberg for helpful comments about dynamic critical phenomena, W.~W. Webb and T.~Baumgart for conversations, and the Institute Curie in Paris, where this work was begun, for hospitality. This work was supported in part by the Harvard Society of Fellows (Y.T.), by National Science Foundation (NSF) Grant DMR-0231631, and through Harvard Materials Research Science and Engineering Laboratory by NSF Grant DMR-0213805.

\appendix

\section*{Supplemental material}

\begin{figure}
\includegraphics[width=0.9\linewidth,clip=]{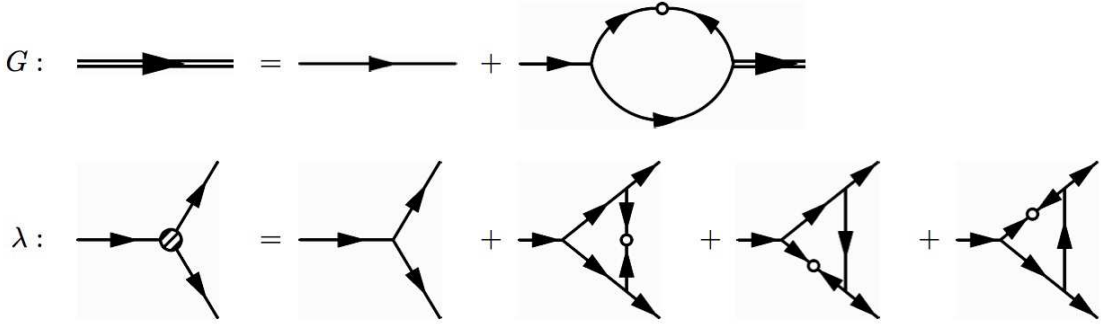}
\caption{Graphical representation of recursion relations for the propagator $G$ (which determines substrate friction $\alpha$ and viscosity $\nu$) and vertex proportional to $\lambda$. Single lines are propagators $G$ with momentum cutoff $\Lambda$, open circles are averaged force correlators (\ref{fdt}) on the momentum shell $\Lambda e^{-l}<q<\Lambda$, and bare vertices are given by $-(i/2)\lambda P_{lij}(\mathbf{k})$ according to the last term in Eq.~(\ref{vl}). Double lines are propagators with scaled cutoff, to order $\lambda^{2}$, and dashed circle is the renormalized vertex, to order $\lambda^{3}$. The intermediate momenta are integrated over the shell $\Lambda e^{-l}<q<\Lambda$.}
\label{vRG}
\end{figure}

Diagrams defining the intermediate recursion relations (before momentum rescaling) for the propagator and vertex following from Eq.~(\ref{vl}) are schematically shown in Fig.~\ref{vRG}. The resulting renormalization-group (RG) flows after momentum rescaling are given by
\begin{align}
\frac{d\alpha(l)}{dl}&=\alpha(l)\frac{d\gamma(l)}{dl}\,,\nonumber\\
\frac{d\nu(l)}{dl}&=\nu(l)\left[\frac{d\gamma(l)}{dl}-2\right]+\frac{\tau}{16\pi}\frac{\lambda(l)^{2}\Lambda^{2}}{\alpha(l)+\nu(l)\Lambda^{2}}\,,\nonumber\\
\frac{d\lambda(l)}{dl}&=\lambda(l)\left[\frac{d\gamma(l)}{dl}-2\right]\,,
\label{rg}
\end{align}
where $\gamma(l)$ is an arbitrary exponent for frequency rescaling. It turns out that only viscosity is modified by reducing the momentum cutoff $\Lambda$ (in particular, the three vertex-renormalization diagrams in Fig.~\ref{vRG} cancel, as in Ref.~\cite{forsterPRA77} which had $\alpha=0$). A convenient choice for $\gamma(l)$ ensures that $\alpha(l)+\nu(l)\Lambda^{2}={\rm const}$.  It is natural to define the renormalized viscosity by the velocity correlator
\begin{equation}
G_{ij}(\mathbf{k},\omega)=\frac{\left\langle v_i(\mathbf{k},\omega),v_j(\mathbf{k}^{\prime},\omega^{\prime})\right\rangle_\mathbf{f}}{(2\pi)^{3}\delta(\mathbf{k}+\mathbf{k}^{\prime})\delta(\omega+\omega^{\prime})}=2\tau P_{ij}(\mathbf{k})\mbox{Re}\frac{1}{-i\omega+\alpha+\nu_{r}(\mathbf{k},\omega)k^{2}}\,.
\label{Gij}
\end{equation}
By the fluctuation-dissipation theorem, this is the same viscosity that enters the full velocity propagator. Upon expressing the velocity $\mathbf{v}$, wave vector $\mathbf{k}$, and frequency $\omega$ in terms of the rescaled quantities, we can calculate $G_{ij}$ if the renormalized parameters flow into a solvable regime. If $\omega=0$ (which will be assumed in the following), RG flows have to be stopped before $e^{l}k=\Lambda$ when the rescaled momentum hits the cutoff. When $k\to0$, $l$ can thus be sent to infinity, and the theory becomes asymptotically free as solving Eqs.~(\ref{rg}) gives $\lambda(l\to\infty)\to0$. Within perturbation theory, small remaining $\lambda$ only results in a $k^{2}$ contribution to the viscosity $\nu_{r}$, which can be safely disregarded in the large-distance limit. We thus have to solve Eqs.~(\ref{rg}) as $l\to\ln(\Lambda/k)$ to find $\alpha(l)$ and $\nu(l)$, which define the linear velocity correlator that can be mapped onto the full correlator (\ref{Gij}) of the original problem. The solution of this problem for the renormalized viscosity can be recast as a solution of the following first-order differential equation:
\begin{equation}
\frac{d\nu_{r}(\bar{k})}{d\bar{k}}=-\frac{\tau}{16\pi}\frac{\bar{k}}{\alpha+\nu_{r}(\bar{k})\bar{k}^{2}}\,,
\label{dn}
\end{equation}
with initial condition $\nu_{r}(\bar{k}=\Lambda)=\nu$. The renormalized viscosity $\nu_{r}(k)$ is given by $\nu_{r}(\bar{k}=k)$. Solving Eq.~(\ref{dn}) results in Eq.~(\ref{nus}) in the limit of $k\to0$.

\begin{figure}
\includegraphics[width=0.7\linewidth,clip=]{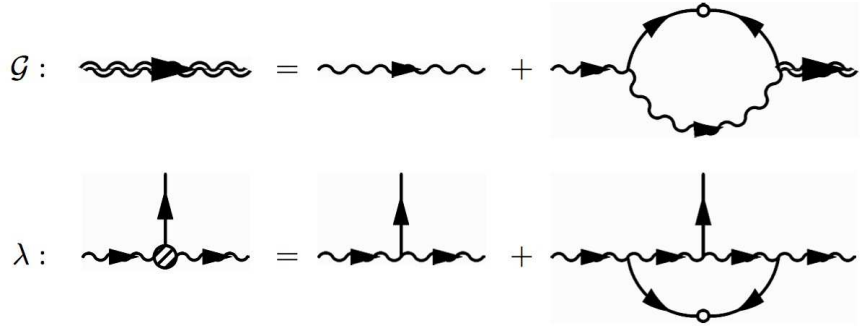}
\caption{Graphical representation of recursion relations for the diffusion propagator $\mathcal{G}$ which determines diffusion coefficient $D$ and vertex proportional to $\lambda$. Single plain lines are propagators $G$, open circles are averaged force correlators (\ref{fdt}) on the momentum shell $\Lambda e^{-l}<q<\Lambda$, and bare vertices are given by $-i\lambda k_i$ according to the last term in Eq.~(\ref{c}). Single wiggly lines are diffusion propagators $\mathcal{G}$ with momentum cutoff $\Lambda$ and double wiggly lines are diffusion propagators with scaled cutoff, to order $\lambda^{2}$. Dashed circle is the renormalized vertex, to order $\lambda^{3}$. The intermediate momenta are integrated over the shell $\Lambda e^{-l}<q<\Lambda$.}
\label{DRG}
\end{figure}

The diagrams representing recursion formulas for the diffusion coefficient $D$ and vertex $\lambda$ corresponding to Eq.~(\ref{c}) are sketched in Fig.~\ref{DRG} \cite{aronovitzPRA84}. As before, nonlinear corrections to the vertex vanish, and we can appropriately rescale $c(\mathbf{k},\omega)$ in the second RG step so that $\lambda$ flows exactly as in Eqs.~(\ref{rg}) [that is why we used the same $\lambda$ in Eqs.~(\ref{ns}) and (\ref{pd})]. We thus have to supplement Eqs.~(\ref{rg}) with only one additional flow equation:
\begin{equation}
\frac{dD(l)}{dl}=D(l)\left[\frac{d\gamma(l)}{dl}-2\right]+\frac{\tau}{4\pi}\frac{\lambda(l)^{2}\Lambda^{2}}{\alpha(l)+\left[\nu(l)+D(l)\right]\Lambda^{2}}\,.
\end{equation}
Upon defining the renormalized diffusion coefficient by
\begin{equation}
\left\langle c(\mathbf{k},\omega)\right\rangle_\mathbf{f}=\frac{c(\mathbf{k},t=0)}{-i\omega+D_r(\mathbf{k},\omega)k^2}
\end{equation}
and setting $\omega=0$, we finally obtain the differential equation similar to Eq.~(\ref{dn}):
\begin{equation}
\frac{dD_{r}(\bar{k})}{d\bar{k}}=-\frac{\tau}{4\pi}\frac{\bar{k}}{\alpha+[\nu_{r}(\bar{k})+D_r(\bar{k})]\bar{k}^{2}}\,,
\label{dD}
\end{equation}
with initial condition $D_r(\bar{k}\sim1/a)=D_0$ (physically meaning that the diffusion-constant flow is ``turned on" at momenta smaller than inverse protein size, assuming $k\ll1/a$). $D_r(\bar{k}=k)$ then gives the renormalized diffusion coefficient $D_r(k)$, which is a measurable quantity describing particle distribution at long times after a certain prepared distribution. Such diffusion coefficients were measured in Ref.~\cite{petersPNAS82} using a fluorescence apparatus. 

We must now solve Eq.~(\ref{dD}) using $\nu_r(\bar{k})$ given by Eq.~(\ref{dn}) as an input (and let us not approximate $\nu_r\approx\nu$ for the moment). This is in general a formidable task analytically. When $\alpha\approx0$, however, equations simplify considerably, and we get for the small-$k$ limit (assuming $a\Lambda\sim1$)
\begin{equation}
D_{r}(k\to0)\approx A\sqrt{\frac{\tau}{8\pi}\ln\left(\frac{\Lambda}{k}\right)}\to A\sqrt{\frac{\tau}{8\pi}\ln\left(\Lambda\delta_r\right)}\,,
\label{Da}
\end{equation}
which is the same as Eq.~(\ref{na}) apart from the numeric prefactor $A=(\sqrt{17}-1)/2$, in agreement with Ref.~\cite{forsterPRA77}. Using the $k=0$ limit of Eqs.~(\ref{na}) and (\ref{Da}), we find that $\nu_r$ and $D_r$ essentially satisfy the relation (\ref{D}), up to a numeric factor of $2/A\approx1.3$, if we set $l^\ast=\delta_r$ and assume $a\Lambda\sim1$. At low temperatures, however, $\nu_r\approx\nu$ and $\nu_r\gg D_r$ [as the diffusion coefficient is in general proportional to $T$ according to the Einstein relation (\ref{E})]. Solving Eq.~(\ref{dD}) then gives Eq.~(\ref{Dr}) for $k\to0$.

\end{document}